\documentclass[smallextended]{svjour3}       
\smartqed  
\usepackage[dvipdfmx]{graphicx}
\usepackage{color}
\usepackage{soul}
\soulregister\cite7
\soulregister\ref7
\soulregister\pageref7
\newcommand{\nasu}[1]{\textcolor{blue}{#1}}
\DeclareRobustCommand{\nchange}[2]{\ifmmode{{\textrm{\setul{}{1pt}\setstcolor{blue}\st{$\displaystyle#1$}}}}\else{{\setul{}{1pt}\setstcolor{blue}\st{#1}}}\fi\ \textcolor{blue}{#2}}

\begin{document}

\title{Magnetic properties in the metallic magnets with large anisotropy}



\author{Yojiro Taguchi \and Joji Nasu \and Akihisa Koga \and Takuya Yoshioka \and Hiroki Tsuchiura}


\institute{Y. Taguchi \and J. Nasu \and A. Koga \at
  Department of Physics, Tokyo Institute of Technology, Meguro, Tokyo 152-8551, Japan
           \and
           T. Yoshioka \and H. Tsuchiura \at
           Department of Applied Physics, Tohoku University, 6-6-05 Aoba, Sendai 980-8579, Japan \\
           ESICMM, National Institute for Materials Science, Tsukuba 305-0047, Japan \\
           Center for Spintronics Research Network, Tohoku University, Sendai 980-8577, Japan
}

\date{Received: date / Accepted: date}

\maketitle

\begin{abstract}
  We study low temperature properties in the metallic magnets,
  considering the itinerant electron mediated ferromagnetism.
  Applying the Monte Carlo simulations to the extended double exchange model,
  we discuss reorientation phase transition and anisotropy field for the metallic magnets.
\end{abstract}

\section{Introduction}
Rare-earth based permanent magnets such as Nd-Fe-B are not only of practical interest
in technology but also provide challenging theoretical problems in fundamental physics~\cite{Herbst,Sagawa,Hirosawa,Yamada1988}.
Their magnetic properties have been extensively examined experimentally.
Theoretical descriptions, however, have not been provided in a consistent way.
The static properties such as magnetocrystalline anisotropy can be described for $T=0$
by using first-principles calculations~\cite{Hummler,Yoshioka1},
and temperature dependences or several
dynamical behavior of the systems have been treated within rather phenomenological way as
classical localized spin models although the systems are metallic~\cite{Yoshioka2,Toga2016,Nishino,Tsuchiura}.
Thus, it is highly desirable to give a consistent description of the dynamical properties
based on a microscopic model consisting of the localized orbitals for $4f$ electrons
interacting with itinerant $3d$ electronic systems.

To discuss the role of the itinerant electrons in the permanent magnets,
we consider the extended double exchange model with the magnetic anisotropy,
which should be one of the simplest models describing magnetic properties in
the metallic permanent rare-earth based magnets Nd-Fe-B.
First, we determine the model parameters by means of the first-principle calculations.
We then study magnetic properties mediated by itinerant electrons at low temperatures
using Monte Carlo simulations.

The paper is organized as follows.
In \S\ref{sec:model}, we introduce the model Hamiltonian relevant for the metallic magnets.
Then, we show the numerical results for the temperature dependent magnetization
and magnetization curves in \S\ref{sec:results}.
A summary is provided in the last section.

\section{Model and Method}\label{sec:model}
In the section, we introduce the model Hamiltonian for the metallic magnets such as Ne-Fe-B.
To capture the essential features,
we here focus on the magnetocrystalline anisotropy and neglect the magnetic moments in the iron ions
although these are expected to play an important role in realizing useful permanent magnets.
Then, the system should be described by the simplified
double-exchange model,
where localized magnetic moments in rare-earth ions correlate with each other
through the conduction electons.
The Hamiltonian is given as,
\begin{eqnarray}
  \label{eq2.1}
  \mathcal{H} = -t \sum_{\langle ij\rangle \sigma}c_{i \sigma}^\dagger c_{j \sigma}
  -2J\sum_{i} {\bf s}_i\cdot {\bf S}_{i}+\sum_i h_{cef}^i
  -\sum_i \left( g {\bf s}_i + g' {\bf S}_i\right)\cdot {\bf H},
\end{eqnarray}
where $c_{i\sigma}$ is an annihilation operator of the conduction electron with spin $\sigma$
at $i$th site,
${\bf s}_i=\frac{1}{2}\sum_{\alpha\beta} c_{i\alpha}^\dag {\bf \hat{\sigma}}_{\alpha\beta} c_{i\beta}$, and
${\bf \hat{\sigma}}$ is the Pauli matrix.
${\bf S}_i$ is the localized $S=9/2$ spin,
which should be the total angular momentum for the neodymium ions.
$t$ is the hopping integral for the itinerant bands,
$J$ is the ferromagnetic exchange coupling
between itinerant and localized electons, and ${\bf H}$ is the magnetic field.
$g(=\nasu{-}2)$ and $g'(=\nasu{-}8/11)$ are the $g$-factors for itinerant electrons and localized spins.
The anisotropy in the rare-earth ions is given as,
\begin{eqnarray}
  h_{cef}^i=\sum_{lm}A_{l}^{m} \langle r^{l}\rangle \>\Theta_{l}O_{l, i}^{m},
\end{eqnarray}
where $A_{l}^{m} \langle r^{l}\rangle$ is the crystalline electric field paramter,
$\Theta_l$ and $O_l^m$ are Steavens factors and operators~\cite{Stevens}, respectively.
\begin{table}[htb]
  \caption{Crystalline electric field parameters $A_{l}^{m}\langle r^{l} \rangle$
  obtained from the first principle calculations for $\mathrm{Nd_2 Fe_{14} B}$
  with $f$ and $g$ inequivalent sites.
  For simplicity, we use its mean values in our Monte Carlo simulations.
  Energy unit in the table is kelvin.}
  \centering
  \begin{tabular}{ccrr} \hline\noalign{\smallskip}
    \ l \ & \ m \ & f site \phantom{8}& g site \phantom{8}\\ \noalign{\smallskip}
    \hline\noalign{\smallskip}
    2 & 0   &  354  & 457 \ \\\noalign{\smallskip}
    2 & -2  & 866  & \ \ \ -262 \ \\ \noalign{\smallskip}
    4 & 0   & -51.2 & -55.6 \ \\ \noalign{\smallskip}
    4 & -2 & \ \ \ -86.5 & 73.8  \ \\ \noalign{\smallskip}
    4 & 4  & -147 & 101  \ \\ \noalign{\smallskip}
    6 & 0  & -2.61 & -2.03 \ \\ \noalign{\smallskip}
    6 & -2 & 4.31  & -3.06 \ \\ \noalign{\smallskip}
    6 & 4   & -27.1 & -18.8 \ \\ \noalign{\smallskip}
    6 & -6 & -4.86 & 4.94  \ \\  \hline
  \end{tabular}


  \label{Alm}
\end{table}
In this study, we do not consider quantum fluctuations in the localized spins, for simplicity.
This assumption may be justified for spins with the large magnetic moment.
In the case, localized spins can be treated as classical vectors.
In the following, the conduction electron density is fixed at the quarter filling.

\begin{figure}[htb]
\includegraphics[width=9cm]{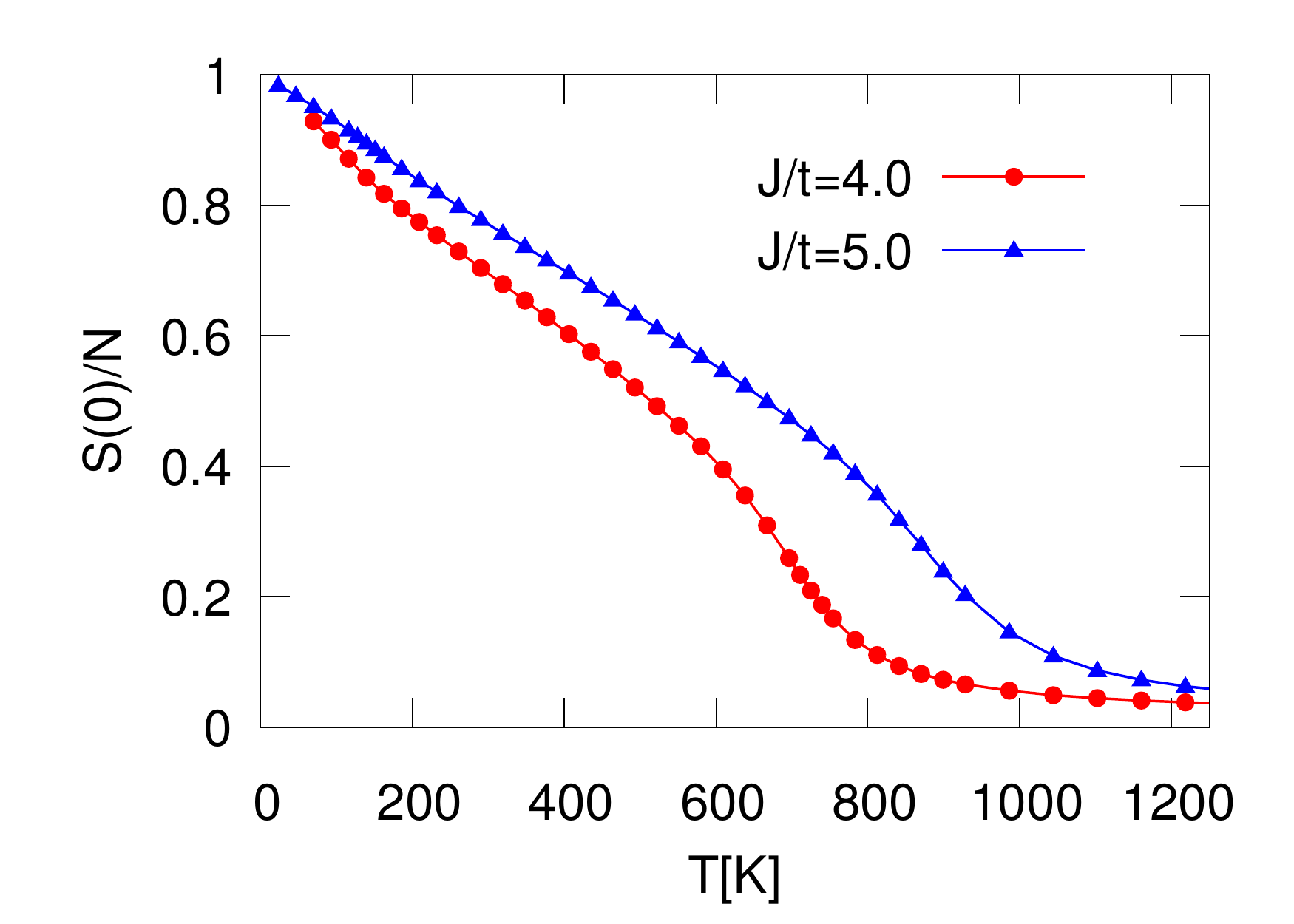}
\caption{
  Squared magnetization $M^2$ as a function of the temperatures
  in the extended double exchange model with $J/t=4$ and $5$.
}
\label{fig1}
\end{figure}

Before discussing magnetic properties in the metallic magnet $\mathrm{Nd_2 Fe_{14} B}$,
we would like to determine the model parameters.
The spin anisotropy parameters, which originate from the crystalline electric fields,
are obtained by means of the first principle calculations for the Nd-Fe-B system.
These are explicitly given in Table~\ref{Alm}.
We fix the hopping integral as $t=1.0\ {\rm eV}$.
The corresponding bandwidth for conduction band $W=6t\sim 6.0\ {\rm eV}$
is a reasonable value of the metallic magnets.
As for the exchange coupling between itinerant and localized electrons,
its magnitude should be determined such that
a calculated transition temperature coincides with the ferromagnetic critical temperature
$T_c\sim 585\ {\rm K}$ for the Nd-Fe-B compound.
In this study, we make use of the Monte Carlo
simulations~\cite{Motome_Furukawa_1999,Motome_Furukawa_2000}
to estimate the appropriate exchange coupling.
Figure~\ref{fig1} shows the temperature dependence of the spin structure factor
at $q=0$, defined by $S(q=0)/N=\sum \langle {\bf S}_i\cdot {\bf S}_j\rangle/N^2$.
In our calculations, the system size is fixed as $N=4^3$ sites.
This system size is not large enough to discuss quantitatively critical phenomena for
the ferromagnetic instability.
Nevertheless, we find that, decreasing temperatures,
the magnetization rapidly increases around a certain temperature,
depending on the ratio $J/t$.
Since the critical temperatures for small clusters are known to be overestimated,
we fix the exchange coupling $J=4t$ in this work.
Using these reasonable parameters,
we examine low temperature properties in the metallic magnets
to discuss the role of the itinerant electrons in the following.

\section{Reorientation transition and anisotropy field} \label{sec:results}

\begin{figure}[htb]
\includegraphics[width=9cm]{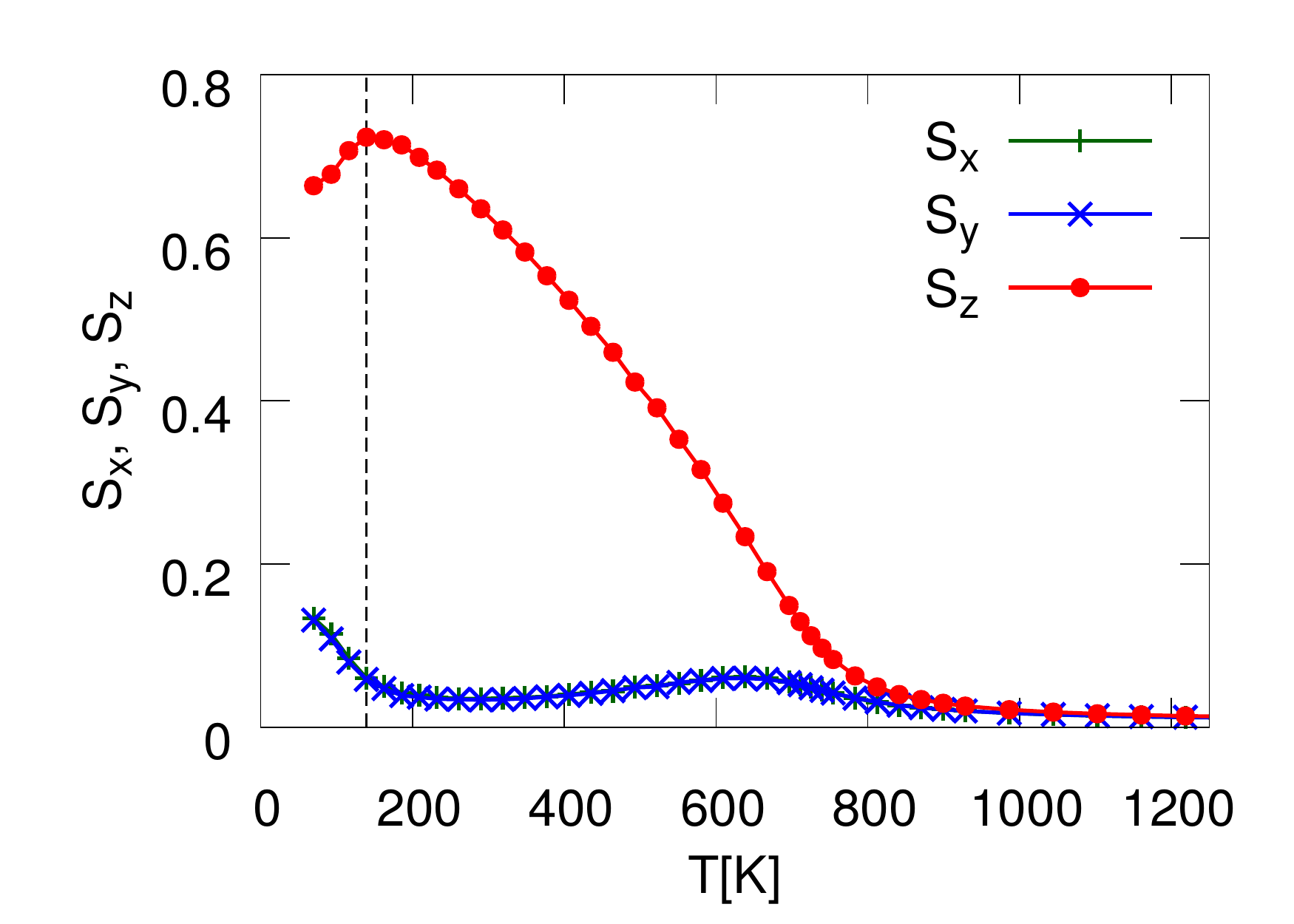}
\caption{
  Magnetization in the $x, y,$ and $z$ directions in the double exchange model
  with $J/t=4$.
}
\label{fig11}
\end{figure}

In this section, we discuss low temperature properties characteristic of
the metallic magnet Nd-Fe-B.
To clarify how the anisotropy in localized spins affects bulk magnetic properties,
we calculate
the $k(=x,y,z)$ component in the spin structure factor
$S_k=\sum \langle S^k_i S^k_j\rangle/N^2$, which are shown in Fig.~\ref{fig11}.
We find that, decreasing temperatures, the magnetization in the $z$ direction
develops below the ferromagnetic transition temperature,
while the magnetizations in $x$ and $y$ directions are almost zero
due to the uniaxial anisotropy of the localized spin.
A remarkable feature is that a cusp singularity appears in both curves around $T\sim 150\ {\rm K}$
although no singularity appears in the total spin structure factor
(see Fig.~\ref{fig1}).
This implies that the anisotropic terms with $m\neq 0$ becomes relevant at $T<150\ {\rm K}$,
which induces canting behavior in the ordered moments.
This is consistent with the fact that
a corresponding reorientation transition has been observed
in the realistic magnets $(T\sim 135\ {\rm K})$~\cite{Kou}.

\begin{figure}[htb]
\includegraphics[width=12cm]{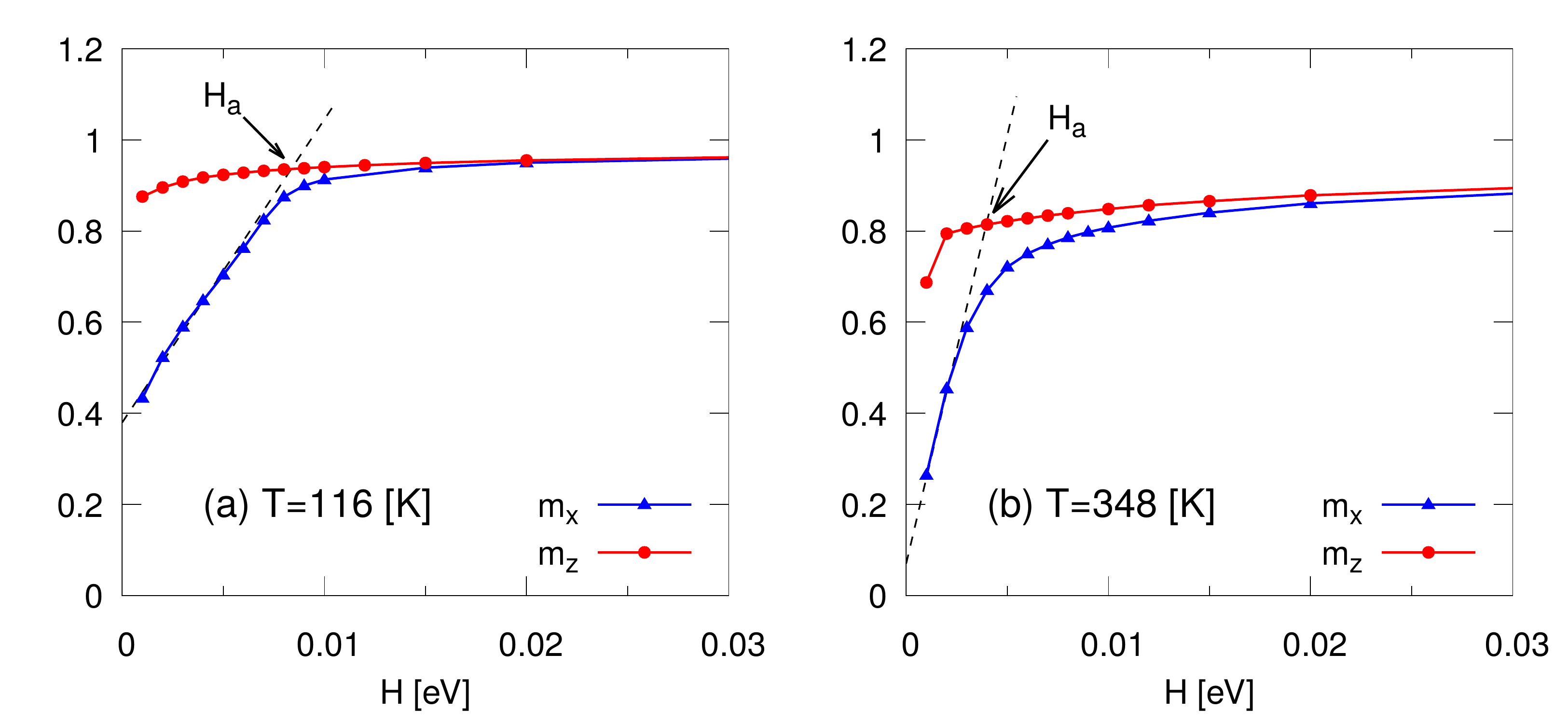}
\caption{
  Magnetization curves $m_x$ and $m_z$ at the temperatures $T=116\ {\rm K}$ and $348\ {\rm K}$
  when the external magnetic field ${\bf H}$
  is applied to the system in the $x$ and $z$ direction, respectively.
  Dashed lines are guides to eyes.
}
\label{fig2}
\end{figure}

We also consider the temperature dependence of the anisotropy field.
The field should be one of the important quantities to evaluate the performance
of the permanent magnets since it is deeply related to its coercivity~\cite{Hirosawa}.
This quantity is determined by the crossing point of two magnetization curves
for different magnetic-field directions, and thereby it
has an advantage to examine in the equilibrium system, in contrast to the coercivity.
Here, we directly calculate the magnetization $m_k(=\sum\langle S_i^k\rangle/N)$ under the magnetic field.
The obtained results are shown in Fig.~\ref{fig2}.
Increasing the magnetic field, both magnetizations approach the satuation values.
However, we could not find that these curves cross each other.
In this study, we roughly deduce the anisotropy field, by estimating
the crossing point between the magnetization curve $m_z$ and the initial slope of $m_x$,
as shown in Fig.~\ref{fig2}.
By performing similar calculations,
we obtain the temperature dependence of the anisotropy field.
\begin{figure}[htb]
\includegraphics[width=9cm]{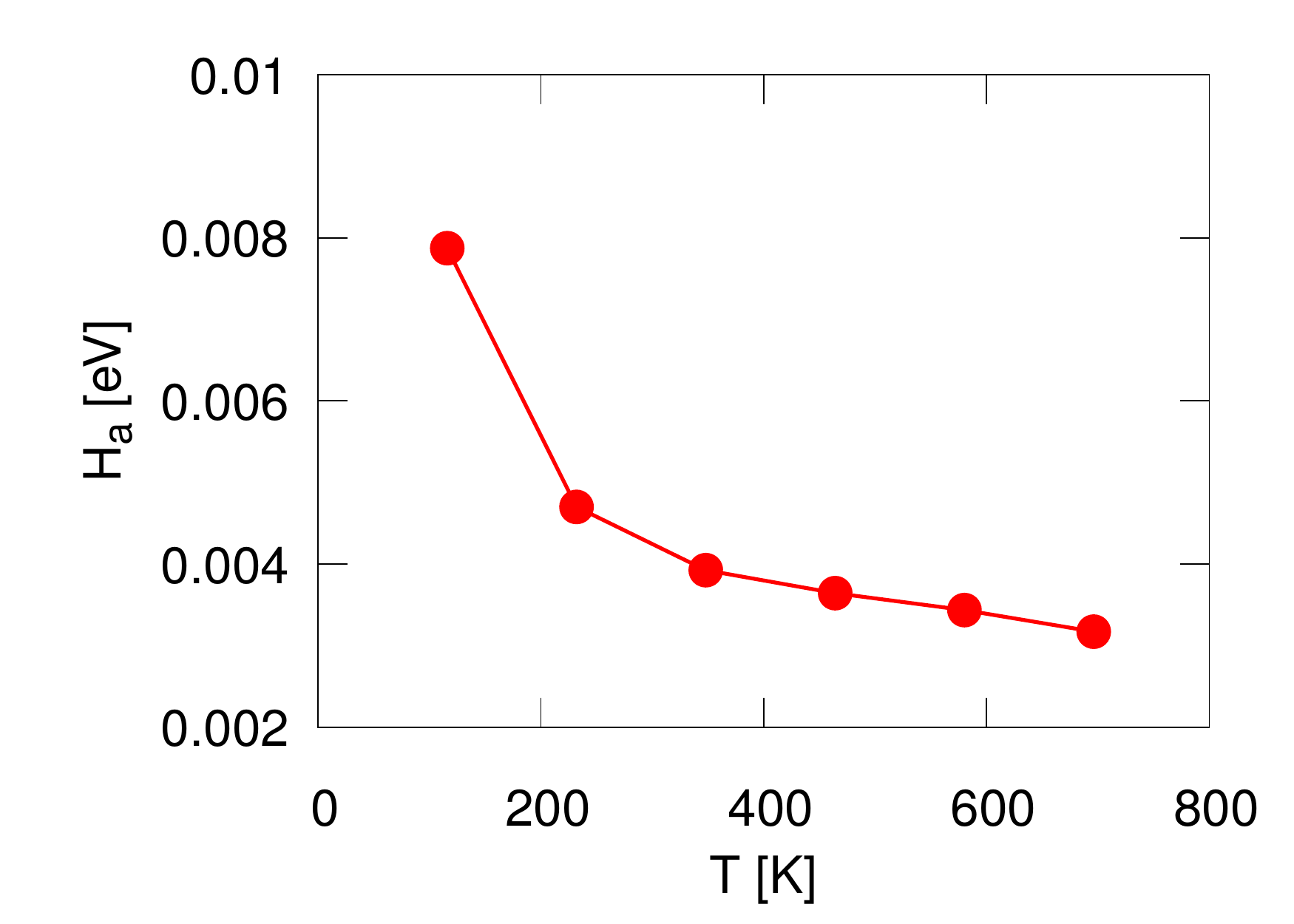}
\caption{
  Anisotropy field $H_a$ as a function of the temperature
  in the double exchange model with $J/t=4$.
}
\label{fig6}
\end{figure}
Figure~\ref{fig6} shows that the anisotropy field monotonically decreases with increase of the temperature.
This is consistent with the experiments for the metallic magnets,
indicating that our microscopic model can well describe these magnetic properties.
It is an interesting problem to clarify how magnetic properties are affected by
the electron density, localized spins in the irons, crystal structure, etc.,
which will be discussed in the future.

\section{Summary}
We have studied the extended double exchange model, which should capture
the essence of the permanent rare-earth based magnet $\rm Nd_2Fe_{14}B$.
To examine the magnetic properties mediated by the itinerant electrons
by means of the Monte Carlo simulations,
we have found that the reorientation transition and anisotropy field are correctly described
in the framework of the simplified microscopic model.

\begin{acknowledgements}
This work is supported by Grant-in-Aid for Scientific Research from
JSPS, KAKENHI Grant Nos. JP17K05536 (A.K.), JP16K17747(J.N.), JP18K04678 (A.K., T.Y. and H.T.).
Parts of the numerical calculations were performed
in the supercomputing systems in ISSP, the University of Tokyo.
The simulations were performed using some of the ALPS
libraries~\cite{alps2}.
\end{acknowledgements}

\bibliographystyle{spphys}       
\providecommand{\url}[1]{{#1}}
\providecommand{\urlprefix}{URL }
\expandafter\ifx\csname urlstyle\endcsname\relax
  \providecommand{\doi}[1]{DOI \discretionary{}{}{}#1}\else
  \providecommand{\doi}{DOI \discretionary{}{}{}\begingroup
  \urlstyle{rm}\Url}\fi

\end{document}